\documentclass{article}

\usepackage{arxiv}

\usepackage[utf8]{inputenc} 
\usepackage[T1]{fontenc}    
\usepackage{hyperref}       
\usepackage{url}            
\usepackage{booktabs}       
\usepackage{amsfonts}       
\usepackage{nicefrac}       
\usepackage{microtype}      
\usepackage{lipsum}
\usepackage{graphicx}
\usepackage{multirow}
\usepackage{epstopdf}

\usepackage{cite}     

\title{Teaching forced damped oscillator using RLC circuit through inverse modeling}

\author{
  K. L. Cristiano \\
  Facultad de Ciencias, Escuela de Física\\
  Universidad Industrial de Santander (UIS)\\
  Colombia, Santander, Bucaramanga \\
  \texttt{karen.cristiano@saberuis.edu.co} \\
  \And
   D. A. Triana \\
	Facultad de Ciencias, Escuela de Física\\
	Universidad Industrial de Santander (UIS)\\
	Colombia, Santander, Bucaramanga \\
   \texttt{ dantrica@saber.uis.edu.co } \\
   \And
   R. Ortiz \\
   Departamento De Matemáticas Y Ciencias Naturales\\
   Universidad Autónoma de Bucaramanga (UNAB)\\
   Colombia, Santander, Bucaramanga\\
   \texttt{rortiz579@unab.edu.co} \\
   \And
      A. F. Estupiñán \\
      Departamento De Matemáticas Y Ciencias Naturales\\
      Universidad Autónoma de Bucaramanga (UNAB)\\
      Colombia, Santander, Bucaramanga\\
      \texttt{aestupinan623@unab.edu.co} \\
}

\begin{document}
\maketitle

\begin{abstract}
Teaching by direct models in science has been weakening the learning process of the students, because the real problems in engineering are not solved by direct models instead commonly they are solve by inverse models.  On the other hand, one of the most relevant topics in the course of waves and
particle physics oriented for the forming engineers; it’s the subject of simple harmonic motion forced
damping, which many physical phenomena can be explained as the quality factor Q and the resonance
frequency of an oscillatory forced system. In order to capture the attention of students and give an
application to this issue. We have developed an experimental setup to take measurements of electric current, voltages from capacitor and inductor for different frequencies and resistances, once the experimental data were collected to study the behavior of the electrical current inside the circuit and find out the RLC parameters with an inverse model. Finally, we want to show the process in detail how parameters of the system (Resistance, Inductance and Capacitance values) are very relevant in this kind of systems, from the results obtained by experimental measurements of voltage, current and angle of phase shift, where this was achieved by implementing an indirect method described in this document, so that can be applied to studies of more complex systems such as a motor where such parameters may be unknown.
\end{abstract}


\section{Introduction}

In many of the current works, experiments of an RLC circuit have been carried out, using numerical and experimental methods to solve the second-order differential equation that governs the behavior of the current along the electric circuit to be studied \cite{r1, r2, r3, r4, r5, r6}.

It is very uncommon to find a research work, in which the students are shown the physical phenomenon of the problem being analyzed, for example the damped effect of the current in a series RLC circuit without previously knowing the values of the resistance, the inductivity of the coil and neither the capacitance of the capacitor used for this experiment.

We authors, we want to present a work in which you can visualize the wave phenomenon of the current, in an RLC circuit, in addition to how from measurements of current and voltage in it, we can see the damping behavior of the system, for which we use a very low resistance value which will allow us to observe the three regions of this system, which are: the inductive region, the resistive region and the capacitive region.

One of the details to highlight in this work, consists of being able to present an RLC circuit from the point of view of the forced oscillatory oscillatory movement, where two solutions are expected, which are: a transient (dominated principally by the factor of the amplitude of voltage of the generator of signals) and another quasistationary (which is presented by the connection of a coil, a resistance and a capacitor), from the two responses obtained (forced and damped) by the physical system studied in this article, we can obtain the values the resistance (R), the inductance (L) and the capacitance (C). For which we have developed a Script in Python, where the results of voltage and current of the system are analyzed principally.

This paper is organized as follows: Section \ref{analítico}, describes the Theoretical study corresponding to the calculates for the RLC circuit. In Section \ref{experimental}, shown the experimental procedure for the data takes. In Section \ref{results} the results concerning to the validate of the experimentall method shown in the previous sections. Finally we present some conclusions about of this work.  

\section{Theoretical study}
\label{analítico}

\begin{figure}[h!]
	\centering
\includegraphics[width=0.7\textwidth]{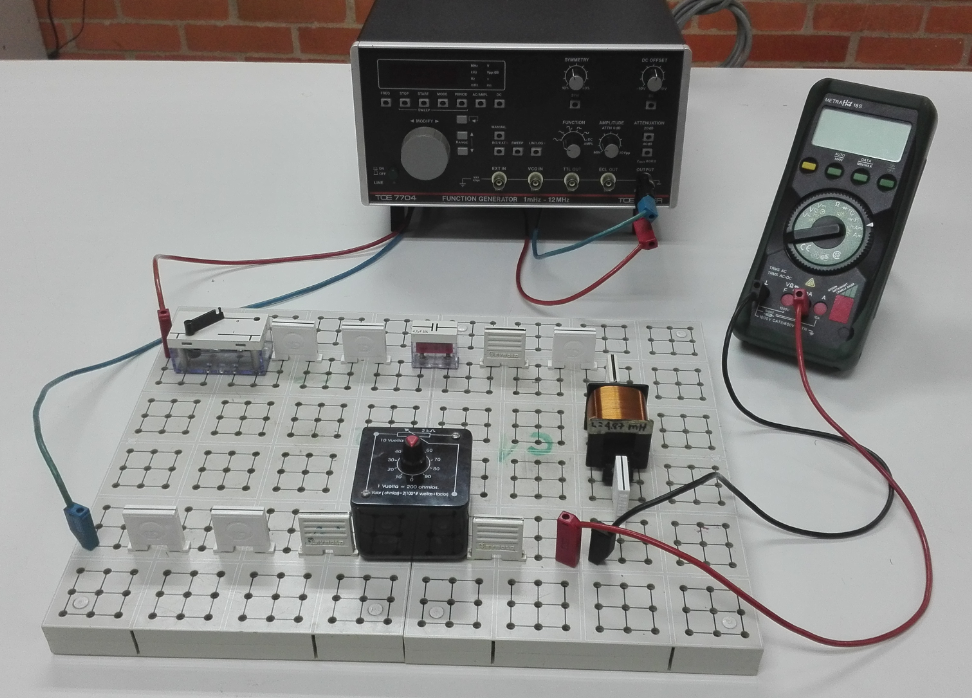}
	\caption{Experimental assembly of the RLC circuit in series.}
	\label{fig_1}
\end{figure}

The physical model of the system under study is shown in Figure \ref{fig_1}. Thissystem consists of an RLC circuit in series.

In order to study this system, we must start with the following equation(Kirchhoff's mesh law) \cite{sears,serway}:

\begin{equation}
V(t) = V_R(t) + V_L(t) + V_c(t)
\label{eq1}
\end{equation}

\begin{equation}
	V_0 \sin(\omega t) = i(t)R + L \frac{di(t)}{dt} + \frac{1}{c} q(t)
	\label{eq2}
\end{equation}

Now using the expression of electric current as a function of charge and time, we have:

\begin{equation}
	i(t) = \frac{dq(t)}{dt}
\label{eq3}
\end{equation}

If we replace the Expression (\ref{eq2}) in Equation (\ref{eq3}), we obtain the following differential equation:

\begin{equation}
	V_0 \sin(\omega t) _= R \left( \frac{dq(t)}{dt} \right) + L \frac{d}{dt} \left( \frac{dq(t)}{dt} \right) + \frac{1}{c} q(t)
	\label{eq4} 
\end{equation}

\begin{equation}
	V_0 \sin(\omega t) _= R \left( \frac{dq(t)}{dt} \right) + L \left( \frac{d^2q(t)}{dt^2} \right) + \frac{1}{c} q(t),
	\label{eq5} 
\end{equation}

where $\omega = 2 \pi f$, being $f$ the temporary frequency of oscillation of the voltage source and by organizing Equation (\ref{eq5}), we have:

\begin{equation}
\frac{V_0}{L} \sin(2 \pi f t) = \frac{R}{L} \left( \frac{dq(t)}{dt} \right) + \left( \frac{d^2q(t)}{dt^2} \right) + \frac{1}{LC} \left( q(t) \right)
	\label{eq6} 	
\end{equation}	

\begin{equation}
\frac{V_0}{L} \sin(2 \pi f t) = \left( \frac{d^2q(t)}{dt^2} \right) + \frac{R}{L} \left( \frac{dq(t)}{dt} \right) + \frac{1}{LC} \left( q(t) \right)
	\label{eq7} 
\end{equation}

From Equation (\ref{eq7}), and using the equation of a forced and damped harmonic oscillator, we can write \cite{zill,tipler}:

\begin{equation}
F_0 \sin(2 \pi f t) = \left( \frac{d^2x(t)}{dt^2} \right) + \gamma \left( \frac{dx(t)}{dt} \right) + \omega^2_0 \left( x(t) \right),
\label{eq8}
\end{equation}
 
where $\gamma$ is the damping factor, $F_0$ is the amplitude of the applied force and $\omega_0$ is the natural frequency of the system. Comparing Equation (\ref{eq7}) with the expression (\ref{eq8}), we can obtain the following expressions:

\begin{equation}
	F_0 = \frac{V_0}{L}, \quad \omega_0 = \frac{1}{\sqrt{LC}}, \quad \gamma = \frac{R}{L}.
	\label{eq9}
\end{equation}

\begin{figure}[h!]
	\centering
	\includegraphics[width=0.5\textwidth]{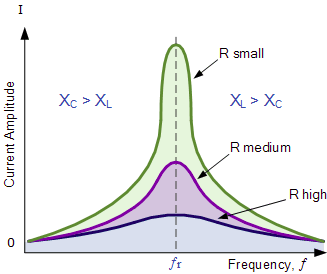}
	\caption{Graph of the current as a function of the frequency, where the expected behavior for three resistances of different values is observed.}
	\label{fig_2}
\end{figure}

Given that the system current $i(t)$ (See the Figure \ref{fig_2}) can be expressed as a function of the voltage $V(t)$ using the following equations,

\begin{equation}
	i(t) = \frac{V(t)}{Z(w)} = \frac{V(t)} {\sqrt{R^2 +  (L\omega - \frac{1}{\omega C})^2}},
	\label{eq10}
\end{equation}

where $z(\omega)$ is the impedance of the circuit, it should be noted that the expression to find the amount of charge stored in the circuit, we can write the harmonic relationship of the charge with the angular frequency of the system $i(t) = I_0 \cos(\omega t + \phi)$, we have:

\begin{equation}
	\frac{dq}{dt} = I_0 \cos(\omega t + \phi)  
	\label{eq11}
\end{equation}

\begin{equation}
	\int dq = \int I_0 \cos(\omega t + \phi) dt
	\label{eq12}
\end{equation}

\begin{equation}
		q(t) =  \frac{I_0}{\omega} \sin(\omega t + \phi) + C.
		\label{eq13}
\end{equation}

If we take into account only the maximum value (amplitude of the function) of the current and the charge we have:

\begin{equation}
	I_0 = \frac{V_0} {\sqrt{R^2 +  (L\omega - \frac{1}{\omega C})^2}}, \quad Q_0 = \frac{V_0} {\omega \sqrt{R^2 +  (L\omega - \frac{1}{\omega C})^2}}.
	\label{eq14}
\end{equation}

\begin{figure}[h!]
	\centering
	\includegraphics[width=0.5\textwidth]{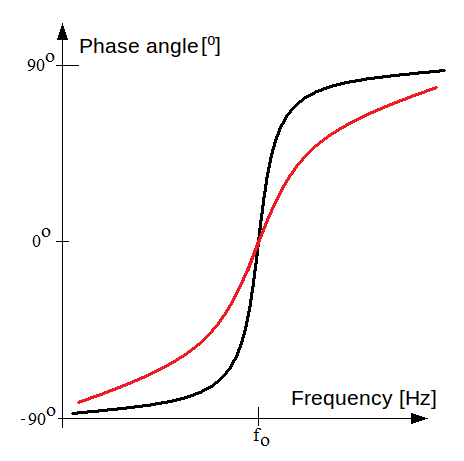}
	\caption{Beta angle behavior $\beta$ for two different resistance values in a series RLC circuit, where the black curve belongs to the value of the lowest value resistance $ R_1 $, so the curve of red color corresponds to the resistance of greater value $ R_2 $ ($R_2> R_1$).}
	\label{fig_3}
\end{figure}

It is also of great importance to define a phase angle $\beta$ of the circuit between the voltage of the coil, that of the capacitor and that of the resistance (See the Figure \ref{fig_3}).

\begin{equation}
	\tan(\beta) = \frac{L\omega - \frac{1}{c\omega}}{R}. 
	\label{eq15}
\end{equation}

One of the most important parameters to be calculated in this type of systems (serial RLC circuit), is the quality factor or merit factor denoted also as factor Q, is defined experimentally as:

\begin{equation}
Q_{exp} = \frac{\omega_{0}}{\Delta f} = \frac{f_{0}}{f_2 - f_1},
	\label{eq16} 	
\end{equation}

Where $f_{0}$ is the resonance frequency of the system, in addition, we can calculate this factor theoretically through the following analytical calculation:

\begin{equation}
Q_{the} = \frac{\omega_{0} L}{R} = \frac{2 \pi f_{0} L}{R}
		\label{eq17}
\end{equation}

\begin{figure}[h!]
	\centering
	\includegraphics[width=0.5\textwidth]{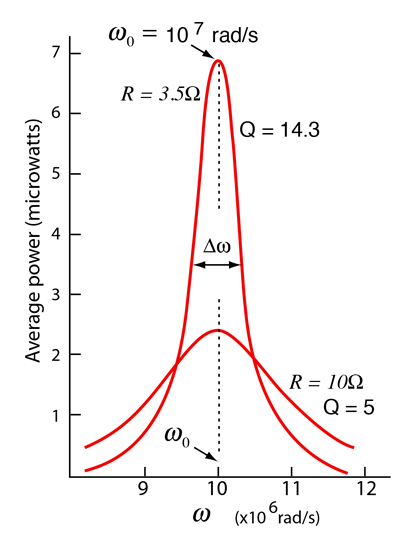}
	\caption{Calculation of the average power of an RLC circuit for the following parameters: $C=2.0$ $nf$, $L=5.0$ $\mu H$ and rms voltage of $5.0$ $mV$.}
	\label{fig_4}
\end{figure}

Finally, a parameter that can give us indications of the behavior of the energy in the circuit is the average power $P_a$ (See the Figure \ref{fig_4}), which is expressed as:

\begin{equation}
P_{a} = V_{a} \cdot I_{a} = V_{0} \cdot I_{0},
			\label{eq18}
\end{equation}

where $I_{a}$ and $V_{a}$ are the average value corresponding of current and voltage. Also $V_{0}$ and $I_{0}$ are the amplitude values of the voltage and current respectively.

\section{Experimental study}
\label{experimental}

We initially carried out the assembly of a series RLC circuit, with the values of resistance R, inductance L and capacitance C supplied. In order for the circuit to behave as forced movement, also add in series an AC source.

Where the procedure to follow is as follows:

\begin{enumerate}
    
    \item For the values of $L = 4,83$ $mH$ and $C = 4,7$ $\mu$F given, calculate the theoretical value of the resonant frequency of the circuit, according to the Equation (\ref{eq9}).
    
    \item Start the data collection with a frequency of 100 Hz and measure the circuit current.
    
    \item Increase the value of the frequency by 100 and measure again, until you find the maximum current of the circuit (in resonance).
    
    \item From the last previous value increase the frequency of the generator in larger intervals (600 Hz) and continue measuring the current until a value close to that measured with the initial frequency of 100 Hz is obtained.
    
    \item  Vary the resistance of the potentiometer and repeat steps 3 to 5.
    
\end{enumerate}

In Table \ref{table_1}, we organize the data taken from the experiment, where we can see that in our case we have a resonance frequency $f_0 = 1044$ $Hz$.

\begin{table}[h!]
\begin{center}
\begin{tabular}{|c|c|c|c|c|}
\hline
\textbf{f [Hz]} & \textbf{I (R = 6 $\Omega$) [mA]} & \textbf{I (R = 20 $\Omega$) [mA]} & \textbf{I (R = 50 $\Omega$) [mA]} & \textbf{I (R = 100 $\Omega$) [mA]} \\ \hline
100 & 5,32 & 5,27 & 5,16 & 4,91 \\ \hline
200 & 10,38 & 10,01 & 9,24 & 7,99 \\ \hline
300 & 14,85 & 13,88 & 12,05 & 9,57 \\ \hline
400 & 18,66 & 16,89 & 13,82 & 10,4 \\ \hline
500 & 21,72 & 19,09 & 14,95 & 10,85 \\ \hline
600 & 24,1 & 20,63 & 15,66 & 11,11 \\ \hline
700 & 25,77 & 21,67 & 16,1 & 11,26 \\ \hline
800 & 26,91 & 22,32 & 16,35 & 11,34 \\ \hline
900 & 27,6 & 22,69 & 16,5 & 11,39 \\ \hline
1000 & 27,82 & 22,85 & 16,55 & 11,4 \\ \hline
\textbf{1044} & \textbf{27,84} & \textbf{22,86} & \textbf{16,6} & \textbf{11,402} \\ \hline
1200 & 27,61 & 22,72 & 16,5 & 11,38 \\ \hline
1800 & 24,29 & 20,72 & 15,69 & 11,1 \\ \hline
2400 & 20,54 & 18,26 & 14,52 & 10,66 \\ \hline
3000 & 17,43 & 15,97 & 13,28 & 10,13 \\ \hline
3600 & 15,04 & 14,07 & 12,12 & 9,58 \\ \hline
4200 & 13,18 & 12,5 & 11,07 & 9,03 \\ \hline
5400 & 10,49 & 11,2 & 10,13 & 8,49 \\ \hline
6000 & 9,5 & 10,12 & 9,32 & 7,99 \\ \hline
6600 & 8,67 & 9,23 & 8,6 & 7,52 \\ \hline
7200 & 7,96 & 8,47 & 8 & 7,1 \\ \hline
8000 & 7,18 & 7,83 & 7,42 & 6,7 \\ \hline
8400 & 6,86 & 7,25 & 6,92 & 6,32 \\ \hline
9000 & 6,41 & 6,76 & 6,49 & 5,98 \\ \hline
9600 & 5,99 & 6,33 & 5,75 & 5,67 \\ \hline
10200 & 5,66 & 5,95 & 5,45 & 5,13 \\ \hline
10860 & 5,32 & 5,27 & 5,16 & 4,91 \\ \hline
\end{tabular}
\vspace{0.2cm}
\caption{Experimental data taken from the RLC circuit current in series, for four different resistance values.}
\label{table_1}
\end{center}
\end{table}

\section{Results}
\label{results}

Once we take the experimental data of the current in the series RLC circuit, using equation (\ref{eq15}), we can obtain the value of the phase angle for the charge $\alpha = - \beta $ and the angular shift of the current $\beta$. In Table \ref{table_2}, we calculate the value of the phase angle $\alpha$ and $\beta$, for the charge and the current respectively, as a function of the temporal oscillation frequency of the source using the electric signal generator.

It is important to note that in Table \ref{table_2}, we have performed the calculation of the $\alpha$ and $\beta$ phase angle with the same frequency intervals for values below the resonant frequency $ f_0 $ (Capacitive regime) and for values above the resonant frequency $ f_0 $ (Inductive regime).

\begin{table}[]
\begin{center}
\begin{tabular}{|c|c|c|c|c|}
\hline
\textbf{f [Hz]} & \textbf{$\alpha$ [$^0$] ($R=6$ $\Omega$)} & \textbf{$\beta$ [$^0$] ($R=6$ $\Omega$)} & \textbf{$\alpha$ [$^0$] ($R=20$ $\Omega$)} & \textbf{$\beta$ [$^0$] ($R=20$ $\Omega$)} \\ \hline
100 & 88,975728634 & -88,975728634 & 86,5894322956 & -86,5894322956 \\ \hline
200 & 87,895055545 & -87,895055545 & 83,0151690125 & -83,0151690125 \\ \hline
300 & 86,69088025 & -86,69088025 & 79,0910790978 & -79,0910790978 \\ \hline
400 & 85,2702154409 & -85,2702154409 & 74,5814380592 & -74,5814380592 \\ \hline
500 & 83,4865468164 & -83,4865468164 & 69,1642654147 & -69,1642654147 \\ \hline
600 & 81,0803124717 & -81,0803124717 & 62,3832531777 & -62,3832531777 \\ \hline
700 & 77,5301965014 & -77,5301965014 & 53,6046128861 & -53,6046128861 \\ \hline
800 & 71,6128821857 & -71,6128821857 & 42,0665903255 & -42,0665903255 \\ \hline
900 & 59,8078349139 & -59,8078349139 & 27,2762553702 & -27,2762553702 \\ \hline
1000 & 30,3629721352 & -30,3629721352 & 9,9678484578 & -9,9678484578 \\ \hline
\textbf{1044} & \textbf{7,1485248931} & \textbf{-7,1485248931} & \textbf{2,1547384631} & \textbf{-2,1547384631} \\ \hline
1200 & -53,8013992912 & 53,8013992912 & -22,2896571884 & 22,2896571884 \\ \hline
1800 & -80,4892669759 & 80,4892669759 & -60,8187979584 & 60,8187979584 \\ \hline
2400 & -84,1662882661 & 84,1662882661 & -71,192734547 & 71,192734547 \\ \hline
3000 & -85,69776159 & 85,69776159 & -75,9224811181 & 75,9224811181 \\ \hline
3600 & -86,5610762758 & 86,5610762758 & -78,6730414637 & 78,6730414637 \\ \hline
4200 & -87,123190111 & 87,123190111 & -80,4908489724 & 80,4908489724 \\ \hline
5400 & -87,8198404866 & 87,8198404866 & -82,7679463229 & 82,7679463229 \\ \hline
6000 & -88,0523884798 & 88,0523884798 & -83,5330624823 & 83,5330624823 \\ \hline
6600 & -88,2390949738 & 88,2390949738 & -84,1488928557 & 84,1488928557 \\ \hline
7200 & -88,3925021602 & 88,3925021602 & -84,6558196597 & 84,6558196597 \\ \hline
8000 & -88,5591992297 & 88,5591992297 & -85,2075262198 & 85,2075262198 \\ \hline
8400 & -88,6300410296 & 88,6300410296 & -85,4422377182 & 85,4422377182 \\ \hline
9000 & -88,7239819496 & 88,7239819496 & -85,7536946762 & 85,7536946762 \\ \hline
9600 & -88,80572858 & 88,80572858 & -86,0249087976 & 86,0249087976 \\ \hline
10200 & -88,8775316953 & 88,8775316953 & -86,2632672432 & 86,2632672432 \\ \hline
10860 & -88,9470770968 & 88,9470770968 & -86,4942434092 & 86,4942434092 \\ \hline
\end{tabular}
\vspace{0.2cm}
\caption{Values of the phase angles $\alpha$ $\beta$ of the charge and current of the RLC circuit as a function of the frequency $f$ for the values of resistance $R = 6$ $\Omega$ and $R = 20$ $\Omega$.}
\label{table_2}
\end{center}
\end{table}

In order to compare the behavior of the phase angle $\alpha$ and $\beta$ for the four different resistance values taken in this experiment ($R = 10, 20, 50$ and $100$ $\Omega$), we show in Table \ref{table_3}, the phase angles for the charge and the current taking into account the highest value of the resistors.

\begin{table}[h!]
\begin{center}
\begin{tabular}{|c|c|c|c|c|}
\hline
\textbf{f [Hz]} & \textbf{$\alpha$ [$^0$] ($R=50$ $\Omega$)} & \textbf{$\beta$ [$^0$] ($R=50$ $\Omega$)} & \textbf{$\alpha$ [$^0$] ($R=100$ $\Omega$)} & \textbf{$\beta$ [$^0$] ($R=100$ $\Omega$)} \\ \hline
100 & 81,5258323067 & -81,5258323067 & 73,4069850003 & -73,4069850003 \\ \hline
200 & 72,9707197615 & -72,9707197615 & 58,5092331715 & -58,5092331715 \\ \hline
300 & 64,2739395152 & -64,2739395152 & 46,0603394561 & -46,0603394561 \\ \hline
400 & 55,4142738385 & -55,4142738385 & 35,948785169 & -35,948785169 \\ \hline
500 & 46,4252981071 & -46,4252981071 & 27,72265128 & -27,72265128 \\ \hline
600 & 37,4009875007 & -37,4009875007 & 20,9214813894 & -20,9214813894 \\ \hline
700 & 28,485951325 & -28,485951325 & 15,179996904 & -15,179996904 \\ \hline
800 & 19,8497996067 & -19,8497996067 & 10,2318390307 & -10,2318390307 \\ \hline
900 & 11,6535938515 & -11,6535938515 & 5,8876870228 & -5,8876870228 \\ \hline
1000 & 4,0212419436 & -4,0212419436 & 2,0130999935 & -2,0130999935 \\ \hline
\textbf{1044} & \textbf{0,8622368479} & \textbf{-0,8622368479} & \textbf{0,431142834} & \textbf{-0,431142834} \\ \hline
1200 & -9,3117916954 & 9,3117916954 & -4,6868441384 & 4,6868441384 \\ \hline
1800 & -35,6127922537 & 35,6127922537 & -19,7041583231 & 19,7041583231 \\ \hline
2400 & -49,5881568393 & 49,5881568393 & -30,4236726718 & 30,4236726718 \\ \hline
3000 & -57,9158297186 & 57,9158297186 & -38,5743814181 & 38,5743814181 \\ \hline
3600 & -63,3995434547 & 63,3995434547 & -44,9557637943 & 44,9557637943 \\ \hline
4200 & -67,2776729403 & 67,2776729403 & -50,0526808255 & 50,0526808255 \\ \hline
5400 & -72,3986576007 & 72,3986576007 & -57,6053246212 & 57,6053246212 \\ \hline
6000 & -74,1784494296 & 74,1784494296 & -60,4573638028 & 60,4573638028 \\ \hline
6600 & -75,630236387 & 75,630236387 & -62,8699498965 & 62,8699498965 \\ \hline
7200 & -76,8372303376 & 76,8372303376 & -64,9333375978 & 64,9333375978 \\ \hline
8000 & -78,1621641571 & 78,1621641571 & -67,2565532532 & 67,2565532532 \\ \hline
8400 & -78,7291642709 & 78,7291642709 & -68,2687197496 & 68,2687197496 \\ \hline
9000 & -79,4844382998 & 79,4844382998 & -69,6330607008 & 69,6330607008 \\ \hline
9600 & -80,144656899 & 80,144656899 & -70,840227315 & 70,840227315 \\ \hline
10200 & -80,7267296339 & 80,7267296339 & -71,9153140726 & 71,9153140726 \\ \hline
10860 & -81,2923208693 & 81,2923208693 & -72,9692652491 & 72,9692652491 \\ \hline
\end{tabular}
\vspace{0.2cm}
\caption{Values of the phase angles $\alpha$ $\beta$ of the charge and current of the RLC circuit as a function of the frequency $f$ for the values of resistance $R = 50$ $\Omega$ and $R = 100$ $\Omega$.}
\label{table_3}
\end{center}
\end{table}

\begin{figure}[h!]
	\centering
\includegraphics[width=0.60\textwidth]{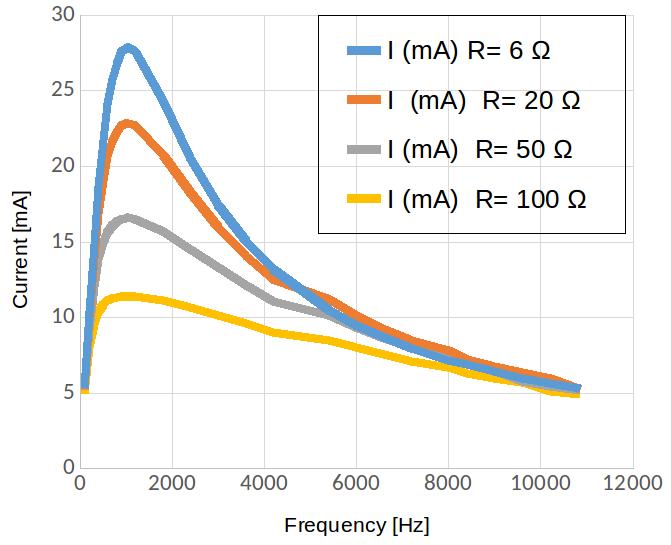}
	\caption{Graph of the current as a function of the frequency of the RLC circuit in series for four different resistance values.}
	\label{fig_5}
\end{figure}

\begin{figure}[h!]
	\centering
\includegraphics[width=0.75\textwidth]{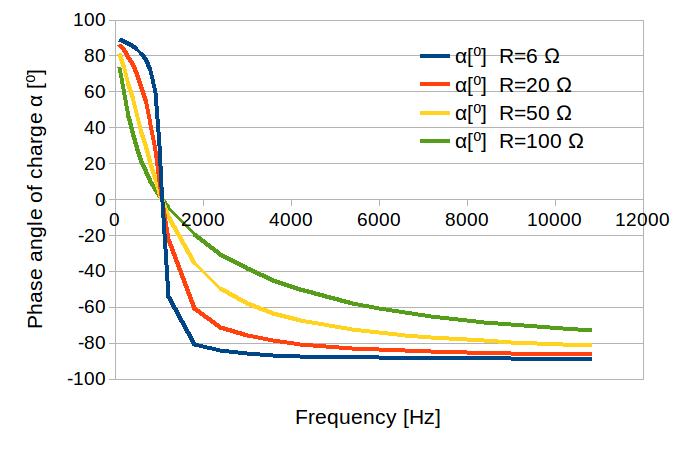}
	\caption{Phase angle of the electric charge, stored in the RLC circuit in series as a function of the frequency for four different resistance values.}
	\label{fig_6}
\end{figure}

\begin{figure}[h!]
	\centering
\includegraphics[width=0.75\textwidth]{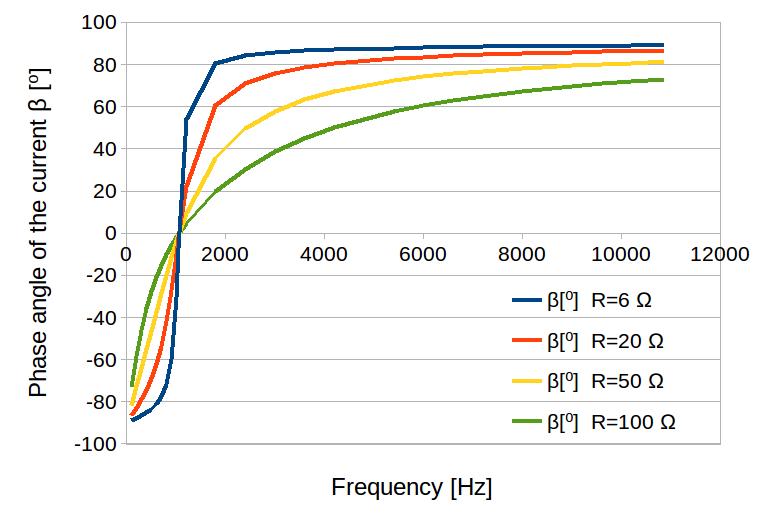}
	\caption{Phase angle of the electric current, circulating in the RLC circuit in series as a function of the frequency for four different resistance values.}
	\label{fig_7}
\end{figure}

In Figure \ref{fig_5}, we show the current curves for the four resistance values that we work in this article, where you can notice the behavior of the current as a function of the frequency, where you can also see the more damped behavior (curve with lower height) to in the case of the higher value resistors ($R = 50$ and $100$ $\Omega$) and, on the contrary, a behavior with a lower damping factor in the case of lower value resistors ($R = 6$ and $20$ $\Omega$), this result corresponds to analytical way with Equation (\ref{eq9}).

In Figure \ref{fig_6} and Figure \ref{fig_7}, we can see the behavior of the phase angle for the electric charge and the current respectively, it should be noted that at a lower resistance value, the graphs show a more curved or smooth behavior and a more linear profile in the case of the highest resistance value.

Also, we show in table \ref{table_4} the results of the experimental and theorical quality factor $Q$ (See the equations (\ref{eq16}) and (\ref{eq17})) for the four resistances, based on the frequency values $f_1$ and $f_2$ using Figure \ref{fig_5}.

\begin{table}[]
\begin{center}
\begin{tabular}{|c|c|c|c|c|}
\hline
\textbf{R [$\Omega$]} & \textbf{$f_1$ [Hz]} & \textbf{$f_2$ [Hz]} & \textbf{$Q$(the)} & \textbf{$Q$(exp)} \\ \hline
6 & 945,14 & 1142,9 & 5,247 & 5,279 \\ \hline
20 & 713,6 & 1374,4 & 1,574 & 1.579 \\ \hline
50 & 220,7 & 1867,34 & 0,629 & 0,634 \\ \hline
100 & 45,7 & 2687,7 & 0,315 & 0,395 \\ \hline
\end{tabular}
\vspace{0.2cm}
\caption{Table of values of the quality factor Q, obtained experimentally and theoretically from equations (\ref{eq16}) and (\ref{eq17}) for the 4 different values of resistance taken in this experiment.}
\end{center}
\label{table_4}
\end{table}

Finally in Table \ref{table_5}, we show the respective experimental errors obtained in this work.

\begin{table}[]
\begin{center}
\begin{tabular}{|c|c|c|}
\hline
\textbf{R [$\Omega$]} & \textbf{Absolute error Q} & \textbf{Percentage relative error Q (\%)} \\ \hline
6 & 0,032 & 0,609 \\ \hline
20 & 0,005 & 0,317 \\ \hline
50 & 0,005 & 0,794 \\ \hline
100 & 0,08 & 25,396 \\ \hline
\end{tabular}
\vspace{0.2cm}
\caption{Table of the calculation of the experimental errors for the quality factor Q for the different resistances taken into account in the experiment.}
\label{table_5}
\end{center}
\end{table}

\section{Conclusions}

In this work the process of implementation of the study of a forced oscillatory movement has been shown in detail, using as an analogy the experimental tool of the assembly of a series RLC circuit, in which it could be seen how the capacitive and inductive region of this case differ at a factor of approximately three orders of magnitude, which is mostly due to the difference in values between the capacitor and the coil used for the development of this article.

With this research, students could be shown a practical example in which the damping factor is involved with the value of the resistance taken in the series RLC circuit, where the relationship between the damping factor and the resistance It is directly proportional, in addition to being able to experimentally observe the value of the resonant frequency in the system for the maximum current amplitude obtained experimentally.

Finally, in this document it was possible to calculate the respective experimental errors obtained for the quality factor Q depending on the value of the resistance taken, where quite acceptable values were obtained.

\section*{Acknowledgments}

The authors would like to thank the Universidad Autónoma de Bucaramanga (UNAB), for lend us the installations and materials for carry to this experiment with which the analytical model presented in this paper could be validated.

\end{document}